# Improved Hodgkin & Huxley-type model for action potentials in squid

P. J. Stiles (1) and C. G. Gray (2)

((1) Department of Molecular Sciences, Macquarie University, Australia

(2) Department of Physics, University of Guelph, Guelph,  Canada)

## Abstract

By extending the crude Goldman-Hodgkin-Katz electrodiffusion model for resting-state membrane potentials in perfused giant axons of squid, we reformulate the Hodgkin-Huxley (HH) phenomenological quantitative model to create a new model which is simpler and based more fundamentally on electrodiffusion principles.  Our dynamical system, like that of HH, behaves as a 4-dimensional resonator exhibiting subthreshold oscillations. The predicted speed of propagating action potentials at 20 degrees Celsius is in good agreement with the HH experimental value at 18.5 degrees Celsius. After the external concentration of calcium ions is reduced, the generation of repetitive rebound action potentials is predicted by our model, in agreement with experiment, when the membrane is stimulated by a brief (0.1 ms) depolarizing current. Unlike the HH model, our model predicts that, in agreement with experiment, prolonged constant-current stimulation does not generate spike trains in perfused axons. Our resonator model predicts rebound spiking following prolonged hyperpolarizing stimulation, observed at 18.5 degrees Celsius by HH but not predicted at this temperature by their quantitative model. Spiking promoted by brief hyperpolarization is also predicted, at room temperature, by our electrodiffusion model, but only at much lower temperatures (*ca*. 6 degrees Celsius)  by the HH model. We discuss qualitatively, more completely than do HH, temperature dependences of the various physical effects which determine resting and action potentials.

## 1. Introduction

In a superb pioneering contribution to computational biology Hodgkin and Huxley (1952) (HH) proposed a quantitative model for the initiation and propagation of action potentials in giant axons of squid. Transient electric-current and steady-state responses of giant axons, initially in their resting states, to a range of clamped membrane potentials were recorded (Hodgkin et al., 1952). These recordings were fitted by HH to a highly empirical dynamical system of four first-order nonlinear differential equations involving the time-dependent membrane depolarization $V(t)$, the sodium-channel activation and inactivation gating functions  $m(t)$ and  $h(t)$, respectively, and a potassium-channel gating function $n(t)$. Each gating function refers to the fraction of open ion gates in a macroscopic sample of the axonal membrane. By treating electrodiffusion in a naturally nonlinear manner (Stiles and Gray, 2019) the present paper avoids the simplifying HH assumption that, in the first of these four equations, ion currents through their respective channels are linear in the membrane potential when sodium and potassium ion gates are fully open; the nonlinearity arising from the dependence of the gating functions $m(t)$ etc on voltage $V$ is also introduced more naturally. Assuming voltage-independent relaxation times, we also simplify the other three HH equations specifying the time-dependence of the three gating functions. This article is both simpler and sounder than an earlier attempt (Perram and Stiles, 2010) to implement a physicochemical electrodiffusive treatment of action potentials. It naturally predicts repetitive firing of action potentials, following a brief initial depolarizing current stimulus, observed in a giant axon of squid after the bathing calcium ion concentration has been reduced (Frankenhaeuser and Hodgkin, 1957; Guttman and Barnhill, 1970; Guttman et al., 1980).

Experiments on squid giant axons (Clay, 1998) indicate that repetitive action potentials, predicted (Ermentrout and Terman, 2010) by the HH quantitative model during prolonged constant electric current stimulation are not observed in perfused axons. Our present theoretical model supports his claim. In contrast, earlier observations (Best, 1979; Chapman, 1980) had suggested that repetitive action potentials occur during prolonged current stimulation of freshly dissected nerve fibers.

The HH quantitative model successfully predicts membrane action potentials in 'anode-break excitation' or 'rebound spiking' experiments at 6.3 °C after a *prolonged* transverse hyperpolarizing electric current stimulus terminates abruptly. The HH model, however, fails to account for the same phenomenon observed at 18.5 °C. Our model accounts for this rebound spiking at both temperatures. The HH quantitative model for rebound spiking also predicts a not well-known normal membrane action potential at 6.3 °C following a *very brief* ( 0.1 ms) hyperpolarizing current stimulus (see Fig. 1A). This phenomenon is fundamental to the notion of facilitatory roles for inhibitory synaptic inputs (Dodla et al., 2006) in neural networks but, it too is not predicted at higher temperatures such as 20 °C by the HH model. As shown in Fig.1B below, it is certainly predicted at this temperature by the simpler resonator version of our present electrodiffusive model. Both models exhibit underdamped subthreshold oscillations, characteristic of resonators (Izhikvich, 2007), as well as resonate-and-fire action potential generation (Izhikevich, 2001, 2007) in response to a very brief superthreshold depolarizing or hyperpolarizing electric perturbation of the resting state.

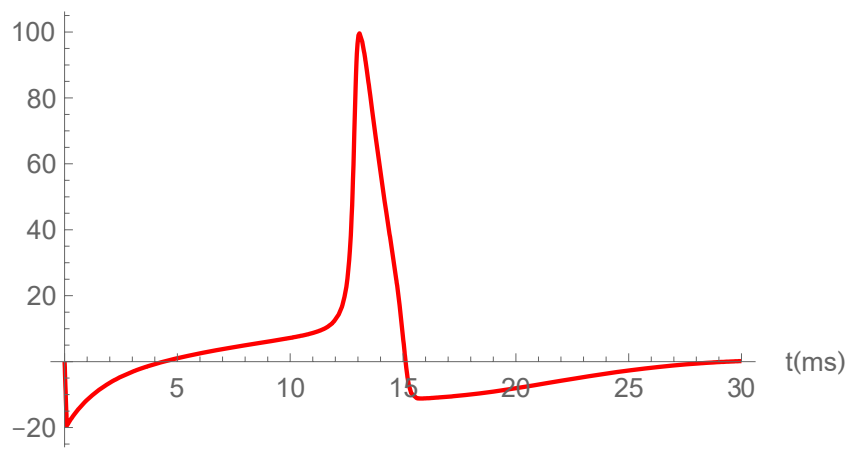

**A** action potential V(t) (mV)

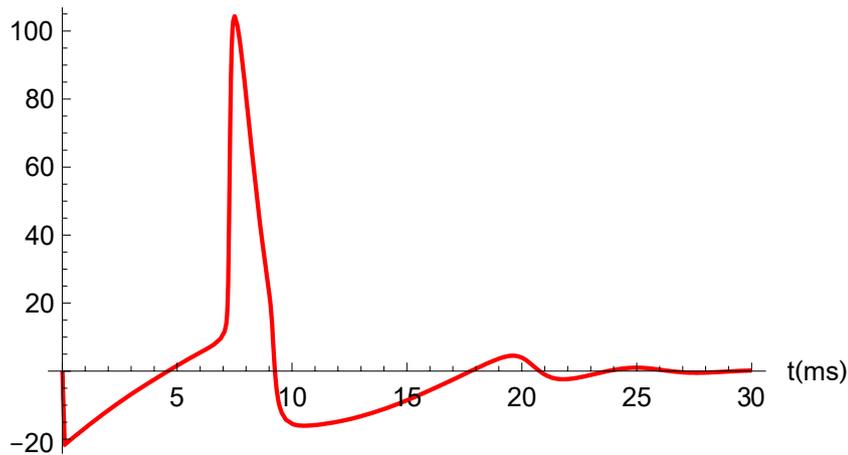

Fig.1. The membrane action potential (A) predicted by the HH quantitative model when a squid axon, initially in its resting state, fires in response to a constant 0.1 ms inhibitory hyperpolarizing current pulse of strength 200 $\mu A\,cm^{-2}$ at 6.3 °C. The corresponding membrane action potential (B) predicted at the higher temperature of 20 °C by our new electrodiffusion model in response to a 0.1 ms hyperpolarizing current stimulus of strength 220 $\mu A\,cm^{-2}$.

    We model axonal spiking in squid more physically than the phenomenological curve-fitting approach to ion conductance adopted by HH by replacing their *ad hoc* $m^3h$ and $n^4$ gating dependencies for sodium- and potassium-ion current densities, respectively, through an extension to the electrodiffusion model of Goldman, Hodgkin and Katz (GHK) (Goldman, 1943; Hodgkin and Katz, 1949a; Perram and Stiles, 2006). We present an alternative mathematical model for action potentials in the spirit of Goldman (1943) who used a constant electric-field approximation to simplify the Nernst-Planck equations for electrodiffusive transmembrane ion currents in the resting state of a nerve fiber. Neither the HH model nor its present refinement consider surface potentials associated with electric double layers (Gray and Stiles, 2018) due to surface charge density at the internal and external aqueous interfaces of the axonal bilayer membrane. Such surface potentials do not affect a true equilibrium resting potential (Gray et al., 2011) but, as discussed later, they influence both membrane steady-state electrodiffusive resting potentials and action potentials. Temporal variations of surface charge densities close to channel orifices might also influence the detailed quantitative description of action potentials when electric currents through ion channels deviate from their resting-state values. In our model for action potentials in a perfused (Baker et al., 1961, 1962) squid axon we imagine that the ubiquitous Na-K-ATPase pumps have been rendered inoperative, so coupling of our passive electrodiffusive fluxes to the active pump fluxes (Armstrong, 2003; Gadsby et al., 2009; Läuger, 1991; Morth et al., 2007; Mullins and Noda, 1963) is artificially suppressed (e.g., by blockers, poisons, or ATP removal). This model can be generalized to re-introduce such bioenergetic coupling which prevents a live squid axon from slowly running down towards a state of thermodynamic equilibrium in which action potentials (spikes) can no longer be generated. Although HH disregarded the GHK model in their celebrated quantitative model for action potentials in squid, a subsequent GHK-like model of Frankenhaeuser and Huxley (1964), for myelinated axons in frogs still retained arbitrary HH-like $m^2h$ and $n^2$ gating dependencies for sodium-

and potassium-ion current densities which our present model, outlined below, avoids. In our electrodiffusion model for channel current-densities, some important nonlinearities are naturally expressed in terms of ion permeabilities which are exponential functions of potentials of mean force (PMFs) $w$ for ions in their specific channels. A PMF is a free energy barrier for ion permeation. As such it depends on various state conditions like temperature. In Section 2.3 we discuss briefly the influence of external $Ca^{2+}$ concentration on the PMF $w_{Na}$. The PMFs $w_{Na}(t)$ and $w_K(t)$ for sodium and potassium ions depend linearly on the gating functions $m(t)$, $h(t)$ and $n(t)$ in our generalized GHK description, and, because the channels are voltage-gated, the gating functions, in turn, depend on the membrane depolarization $V(t)$ which is time-dependent when the membrane is not in a steady-state.

As far as we are aware, ours is the first four-variable model to improve and simultaneously simplify the HH treatment for squid. There have been numerous two-variable models proposed (Abbott and Kepler, 1990; Ermentrout and Terman, 2010; Izhikevich, 2007; Kepler et al., 1992; Krinski and Kokoz, 1973; Nossal and Lecar, 1991; Plant, 1981) which simplify visualization of the complicated nonlinear dynamics in a four-dimensional phase space. Various two-variable models can be derived approximately from four-variable models by assuming the $m(t)$-gate is fast compared to the other two gates, and such reductions render the dynamics more visualizable in a two-dimensional phase plane. Some comparisons of the quantitative predictions of some reduced models with the HH model have been performed (Izhikevich, 2007; Lecar and Nossal, 1971; Kepler et al., 1992; Rinzel, 1978; Rinzel, 1985) and it would be of interest to see similar comparisons with other reduced models, in order to assess both quantitative and qualitative errors in the reduced models.

Consider the mean or macroscopic membrane current density $i$ (in amperes per $m^2$) in the transverse direction, denoted $x$, due to an ion species of charge $q$. Each species of ion is assumed to permeate through the membrane of thickness $L$ *via* ion channels dedicated to that species. If $e$ represents the proton charge then $q = +e$ for sodium and potassium ions and $q = -e$ for chloride ions. Our idealized ion channels for each ionic species are cylindrical pores with symmetry axes orthogonal to the membrane surfaces. Such channels are assumed to account for a surface fraction $f$ of the membrane area. We assume model channels perfectly selective for ions of a given species; this is a reasonable first approximation since under normal conditions (Hille, 2001), Na channels are selective for $Na^+$ ions over $K^+$ ions by 12:1, K channels are very selective for $K^+$ ions over $Na^+$ ions by orders of magnitude and these channels are all very selective with respect to charge sign (perm-selective). The Nernst-Planck-Kramers equation (Perram and Stiles, 2010) for electrodiffusion of ions, with an average intra-channel number-density $c(x,t)$ at an axial position $x$ and time $t$, tells us that the membrane current-density $i(x,t)$ for a particular ion passing through its specific channels is

$$i(x,t) = -q f D \left( \partial c(x,t)/\partial x + \beta c(x,t) \, \partial w_{tot}(x,t)/\partial x \right) \quad , \qquad (1)$$

where $D$, assumed constant, is the ion diffusion coefficient in its specific channel, $w_{tot}(x,t)$ is the total PMF of the ion of interest in its specific channel, and the coefficient $\beta = 1/k_B T$ is defined in terms of the Boltzmann constant $k_B$ and the absolute temperature $T$. The terms proportional to $\partial c/\partial x$ and $\partial w_{tot}/\partial x$ in (1) are called "diffusion" and "drift" terms respectively. We regard the PMF $w_{tot}(x,t)$ of a channel ion as the sum of an electrostatic contribution $q\phi(x,t)$ from the electrostatic potential $\phi(x,t)$ in the essentially one-dimensional ion channel and a second contribution $w(x,t)$ from all other forces between the permeant ion and its channel, including the electric polarization energy. Following Goldman (1943) we replace the local axial electric field $E(x,t) \equiv E_x(x,t) = -\partial\phi(x,t)/\partial x$ in a channel by a spatially constant value, its spatially averaged value between the internal cytoplasmic membrane interface $x = 0$ and the external interface at $x = L$, and write the electric field, common to each ion channel, as $E(x,t) = V_m(t)/L \equiv E(t)$. In defining $\phi(x)$ we have neglected negative membrane surface charges and their concomitant double layers. One of the effects of the double layers is to extend the non-constant regions of $\phi(x)$ beyond the membrane interior $x = (0, L)$; $\phi(x)$ is also variable in the axonal interior and external regions, $(-l_{DL}^{int}, 0)$ and $(L, L + l_{DL}^{ext})$ respectively, where $l_{DL}^{int}$ and $l_{DL}^{ext}$ are the effective interior and exterior double-layer widths, respectively, where $l_{DL} \sim$ a few $l_D$ and $l_D$ is the Debye screening length. If we were to include electric double layers in an electrodiffusion model (which we do not attempt in this paper), the value of the effective Goldman constant electric field would change from $V_m/L$ to $V_m/L'$ where $L' = L + l_{DL}^{int} + l_{DL}^{ext}$. For the resting state of the axonal membrane of squid the membrane potential $V_m = V_{rest}$ is negative and takes an experimental value (Moore and Cole, 1960) close to $-70$ mV at $20°C$ for the normal resting state, i.e., with sodium-potassium pumps operating. As we discuss below, the resting potential in the perfused state with these pumps off (Baker et al., 1961; Mullins and Noda, 1963) is closer to $-68$ mV at $20°C$. The depolarization $V(t)$ at time $t$ is defined by $V(t) = V_m(t) - V_{rest}$. A neural membrane is said to be depolarized if $V(t) > 0$ and to be hyperpolarized when $V(t) < 0$. Using the constant-field approximation (Goldman, 1943) we find that the quasi-steady-state solution to (1), regarded as an ordinary first-order differential equation with fixed boundary conditions $c(0,t) = c^{int}$ at $x = 0$ and $c(L,t) = c^{ext}$ at $x = L$, takes the form (Perram and Stiles, 2010)

$$i(t) = \frac{qfD\left(c^{int} - c^{ext}\exp(-\beta q V_m(t))\right)}{\int_0^L \exp\left(\beta\left(w(x,t) - q(x/L)V_m(t)\right)\right)dx}, \qquad (2)$$

where, with our choice of coordinate system, an outward current is positive. Note that $i = 0$ when $V_m$ assumes the value $(k_B T/q)\ln(c^{ext}/c^{int})$, the membrane Nernst or equilibrium potential for the particular ionic species.

We adopt the simplest possible model for the PMF and assume $w(x,t)$ inside a channel is spatially constant so that $w(x,t) = w(t)$. This PMF $w$ is assumed to vanish outside the channel. For chloride channels, we assume no gating and therefore take $w_{Cl}$ to be independent of time. For potassium and sodium channels, the spatially constant value of each PMF depends on whether the channel is open, closed, or transitioning between these states. Temporal gating of potassium channels and sodium channels leads to PMFs $w_K(t)$ and $w_{Na}(t)$ respectively. In the channels of the monovalent cations sodium and potassium, $q = e$, so that (2) reduces to

$$i(t) = \frac{\beta e^2 P(t) V_m(t) \left(c^{int} - c^{ext} \exp(-\beta e V_m(t))\right)}{1 - \exp(-\beta e V_m(t))} \quad , \tag{3}$$

a simple extension of the traditional resting-state version of the GHK equation for ionic current densities, in which both the resting membrane potential $V_m = V_{rest}$ and ion permeabilities $P$ of the original GHK model have been replaced by their dynamic generalizations $V_m(t) = V_{rest} + V(t)$ and $P(t)$ which vary on a millisecond timescale. The latter take the form

$$P_{Na}(t) = (f_{Na} D_{Na} / L) \exp(-\beta w_{Na}(t)) \quad , \tag{4a}$$

$$P_K(t) = (f_K D_K / L) \exp(-\beta w_K(t)) \quad . \tag{4b}$$

In the absence of chloride-channel gating, the chloride ion permeability takes the static form

$$P_{Cl} = (f_{Cl} D_{Cl} / L) \exp(-\beta w_{Cl}) \quad . \tag{4c}$$

In (4a) and (4b) the cation PMFs $w_{Na}$ and $w_K$ depend on the action potential $V(t)$ through the time-dependent gating functions $m(t)$, $h(t)$ and $n(t)$. It follows that the cation permeabilities $P_{Na}$ and $P_K$ are also functions of time. For potassium ion channels we write the PMF as

$$w_K(t) = n(t) w_K^O + (1 - n(t)) w_K^C \quad , \tag{5}$$

where $n(t)$ is the fraction of open potassium gates at a time $t$, $w_K^O$ is the PMF of a K channel with a fully open ($n = 1$) gate, and $w_K^C$ is the K channel PMF with a completely closed ($n = 0$) gate. As we discuss below, the gating functions such as $n$ depend on the membrane potential $V_m$. Thus $n = n(V_m)$ and in the time-dependent case $n(t)$ can also be denoted as $n(V_m(t))$ or $n(V(t))$. The PMFs in (5) are measured relative to that of a potassium ion in the external bulk electrolyte. Similarly, for sodium-ion channels we define

$$w_{Na}(t) = m(t) w_{Na}^O + (1 - m(t)) w_{Na}^C + h(t) w_{Na}^{\dagger O} + (1 - h(t)) w_{Na}^{\dagger C}, \tag{6}$$

where $m(t)$ refers to the fraction of open activation gates, and $h(t)$ to the fraction of open inactivation gates. The four Na-channel PMF barriers $w_{Na}^C$, $w_{Na}^O$, $w_{Na}^{\dagger C}$ and $w_{Na}^{\dagger O}$ refer, respectively, to the completely closed ($m = 0$) and completely open ($m = 1$) states of the sodium activation gate, and the completely closed ($h = 0$) and completely open ($h = 1$) states of the sodium inactivation gate. The analysis of HH suggests that the resting-state sodium- and potassium-channel gating functions take the temperature-independent values $m = m_\infty(V = 0) \approx 0.053$, $h = h_\infty(V = 0) \approx 0.60$ and $n = n_\infty(V = 0) \approx 0.32$ with the Na-K-ATPase pumps switched on. When the excitable membrane is initially in its perfused resting state (Baker et al., 1961, 1962) a somewhat different picture emerges from our present formalism. In the following section we discuss our version of gating kinetics inspired by, but simpler than the first-order scheme originally proposed by HH. An action potential $V(t)$ is generated when gating kinetics causes the potentials of mean force $w_{Na}(t)$ and $w_K(t)$ for the cation channels to be appropriately synchronized.

## 2. Resting and action potentials

Magnetic fields due to intra-membrane currents are normally negligible because the speeds of ions traversing the axonal membrane are so low. Thus, in the dissipative resting state of the membrane, the sum of the electrodiffusive steady-state chloride, sodium and potassium ion current densities is zero, as demanded by the magnetic-field free version of the Maxwell equation, $\partial D_x / \partial t = -i_{tot} = -(i_{Cl} + i_{Na} + i_K) = 0$ for the temporal derivative of the *x*-component, perpendicular to the membrane surface, of the electric displacement field $\boldsymbol{D}$. From the resting-state version of (3), we then find that the GHK resting potential (Hodgkin and Katz, 1949a) is given by the well-known expression

$$V_{rest} = \frac{k_B T}{e} \ln \frac{P_{Na} c_{Na}^{ext} + P_K c_K^{ext} + P_{Cl} c_{Cl}^{int}}{P_{Na} c_{Na}^{int} + P_K c_K^{int} + P_{Cl} c_{Cl}^{ext}} \quad . \tag{7}$$

For our model, the constant resting-state permeabilities in (7) at the time $t = 0$ are given by

$$P_{Na} = (f_{Na} D_{Na} / L) \exp(-\beta w_{Na}(0)) \quad , \tag{8a}$$

$$P_K = (f_K D_K / L) \exp(-\beta w_K(0)) \quad , \tag{8b}$$

and

$$P_{Cl} = (f_{Cl} D_{Cl} / L) \exp(-\beta w_{Cl}) \quad . \tag{8c}$$

When transmembrane ion currents are time-dependent, the membrane potential $V_m(t) = V_{rest} + V(t)$ also becomes a function of time. Depolarization of a small region of the membrane by an external electric stimulus (voltage or current) with a magnitude exceeding some threshold value (not always precisely the same), produces a characteristic temporal response $V(t)$ of the local membrane potential, known as the membrane action potential, or space-clamped, or stationary, or non-propagating action potential. We might expect the action potential threshold to occur in the region where the sodium activation gating variable $m(V)$ begins to rise most steeply from near zero towards unity ($V \sim 6$ mV, as in Fig. 2 below). Our analysis, indeed, reveals a very narrow threshold beginning at 6.551 mV for our resonator model of the squid axon. We refer to Izhikevich (2007) for a detailed discussion of distinctions between "resonator" and "integrator" neurons. The question of whether such a threshold for the initiation of an isolated action potential is mathematically sharp has long been debated ( Cole et al., 1970; Fitzhugh, 1955; Hodgkin and Huxley, 1952; Izhikevich, 2007). Theoretical models for resonators (Izhikevich, 2007), such as that of HH suggest that at least for squid, an action potential can be initiated for depolarizations in an extremely narrow distribution of threshold values, sometimes called a fuzzy threshold. For integrators with a second fixed point, unstable and near the resting potential (Izhikevich, 2007; Perram and Stiles, 2010), one might expect an even sharper threshold. In addition to dynamical mechanisms leading to fuzzy thresholds, thermal fluctuations of current, voltage and conductance can also lead to such effects (Lecar and Nossal, 1971; Chow and White, 1996). The stimulus for action potential generation frequently takes the form of an initially applied depolarizing voltage shock, $V(0) = V_{stim}$, or of a transverse depolarizing electric current-density, $i_{stim}(t)$ injected briefly into the membrane. Like the electrodiffusive resting potential, the space-clamped membrane action potential is also predicted from the magnetic-field-free Maxwell equation (in SI)

$$\partial D_x / \partial t = (\varepsilon_m \varepsilon_0 / L) \partial V_m / \partial t = -i_{tot}(t) \quad , \tag{9a}$$

or equivalently

$$C_m \dot{V} = -\left(i_{stim}(t) + i_{Cl}(t) + i_{Na}(t) + i_K(t)\right) \quad , \tag{9b}$$

where $C_m = \varepsilon_m \varepsilon_0 / L$ represents the membrane capacitance per unit area, $\varepsilon_m$ the relative membrane permittivity (dielectric constant), $\varepsilon_0$ the permittivity of free space, and $\dot{V} = dV/dt$. The symbol $i_{tot}(t)$ in (9a) refers to the sum of any stimulating current-density, $i_{stim}(t)$, and the current densities of all other permeant ions. Using (3) - (6) we can express all time-dependent ionic current densities in (9b) in terms of the resting potential, the time-dependent membrane depolarization $V(t)$, the sodium-channel gating functions $m(t)$ and $h(t)$, and the potassium-channel gating function $n(t)$. HH assumed that sodium-channel gating was determined exclusively by the membrane potential through the first-order relaxation equations

$$\dot{m} = (m_{ss}(V) - m(t))/\tau_m(V) \quad , \tag{10a}$$

and

$$\dot{h} = (h_{ss}(V) - h(t))/\tau_h(V) \quad , \tag{10b}$$

where $m_{ss}(V)$ and $\tau_m(V)$, together with $h_{ss}(V)$ and $\tau_h(V)$, refer to values of steady-state gating functions and relaxation times, respectively, for the activation and inactivation gates when the membrane depolarization is $V$. Note that our $m_{ss}(V)$ etc, are denoted $m_\infty(V)$, etc, by HH. Subsequent investigations (Bezanilla and Armstrong, 1977; Patlak, 1991; Vandenberg and Bezanilla, 1991) indicated that sodium inactivation gating in squid is directly related to the activation gating but only indirectly to the membrane potential. Although the extent of direct voltage involvement in sodium channel inactivation gating for squid remains somewhat controversial (Vandenberg and Bezanilla, 1991), we have suggested (Perram and Stiles, 2010) that it may be useful to replace $h_{ss}(V)$ in (10b) by $h_{ss}(m)$ where $m$ is a function of $V$. If we further assume that each gating relaxation time can be regarded as constant (Perram and Stiles, 2010), equations (10a) and (10b) can be replaced by their simpler versions

$$\dot{m} = (m_{ss}(V) - m(t))/\tau_m \quad , \tag{11a}$$

and

$$\dot{h} = \left(h_{ss}(m) - h(t)\right)/\tau_h \quad , \tag{11b}$$

respectively. We also replace the HH relaxation equation $\dot{n} = \left(n_{ss}(V) - n(t)\right)/\tau_n(V)$ for potassium-channel gating by its simpler counterpart

$$\dot{n} = \left(n_{ss}(V) - n(t)\right)/\tau_n \quad , \tag{11c}$$

where $\tau_n$ is assumed constant.

When the stimulus for the generation of an action potential takes the form of a voltage shock, the stimulating current-density $i_{stim}(t)$ can be set equal to zero. Equation (9b) for the temporal derivative of the membrane depolarization with ion currents defined by (3) – (6), together with the differential equations (11) for ion gating, constitute an autonomous dynamical system of four first-order nonlinear differential equations for the four dynamical variables $V(t)$, $m(t)$, $h(t)$ and $n(t)$. This dynamical system is then completely defined by assigning algebraic forms for the steady-state gating functions $m_{ss}(V)$, $h_{ss}(m)$ and $n_{ss}(V)$, and specifying the seven PMF barriers $w_{Na}^O$, $w_{Na}^C$, $w_{Na}^{\dagger O}$, $w_{Na}^{\dagger C}$, $w_K^O$, $w_K^C$ and $w_{Cl}$, together with initial conditions on the four dynamical variables. For voltage-shock stimulation of an axon in its resting state, these initial conditions take the form $V(0) = V_{stim}$, $m(0) = m_{rest}$, $h(0) = h_{rest}$ and $n(0) = n_{rest}$ where $m_{rest} = m_{ss}(V = 0)$, etc. We assume that steady-state sodium gating functions take the sigmoidal forms

$$m_{ss}(V) = \left(1 + \tanh\left(s_m(V - V_T)\right)\right)/2 \quad , \tag{12a}$$

$$h_{ss}(m) = \left(1 - \tanh\left(s_h(m - m_T)\right)\right)/2 \quad , \tag{12b}$$

where $m_T$ and $V_T$ are temperature-dependent parameters, with $m_{T=20°C} = 0.26$ and $V_{T=20°C} = 12$ mV. Our steady-state potassium-channel gating function is assumed to be

$$n_{ss}(V) = \left(1 + \tanh\left(s_n V\right)\right)/2 \quad , \tag{12c}$$

with $V$ in mV in (12a,c). These curves sketched in Fig. 2 for $20°C$, with steepness parameters $s_m = 0.16 \, (mV)^{-1}$, $s_h = 11$ and $s_n = 0.15 \, (mV)^{-1}$, ensure that in the perfused resting state of the squid membrane at $t = 0$, $m(0) \approx 0.021$ so that the sodium $m$ gates are only slightly open while $h(0) \approx 0.995$, ensuring that the sodium $h$ gates are almost completely open. Our curve (12c) implies that for potassium gating $n(0) = 0.5$, so in the resting state of a perfused axon with $V = 0$ our potassium channel gate is halfway open, rather than about 32% open as in the HH model described above. These partially open gates in the pumps-off resting state, together with the finite sizes of the PMF free energy barriers are responsible for the resting potential of this state; due to electrodiffusion the resting state is a zero-net-current dynamic steady state in which the inward sodium current is balanced by outward potassium current (for simplicity we ignore the small chloride ion current here). In comparison, in the pumps-on resting state the electrodiffusion sodium/potassium current ratio is 3/2, which balances the pump-driven sodium(outward)/potassium(inward) current ratio of 3/2.

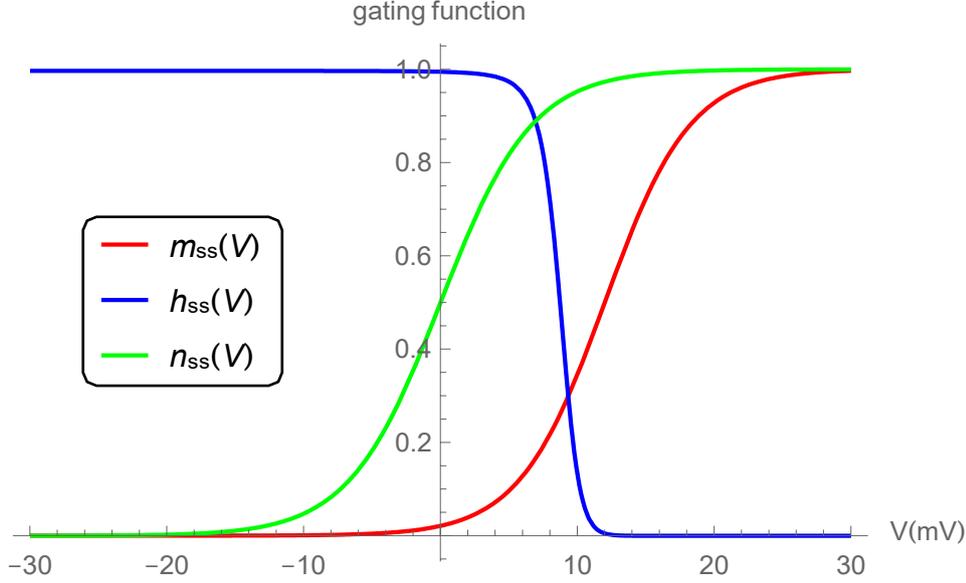

Fig. 2. Our steady-state sodium ion activation gating function $m_{ss}(V)$ (red), sodium-ion inactivation gating function $h_{ss}(V) \equiv h_{ss}(m = m_{ss}(V))$ (blue), and potassium ion gating function $n_{ss}(V)$ (green) plotted against membrane depolarization $V$ at $20°C$.

Other parameters were selected to ensure that stationary membrane action potentials $V(t)$ (see Fig. 4) have temporal profiles similar to those recorded by HH. Our selected values for the seven PMF barriers $w$, three cation gating relaxations times $\tau$, and channel membrane area-fractions $f$ used this criterion. We also used the additional requirement that voltage-clamp electric current transients and steady states should, at least qualitatively, resemble those measured by HH. In Table 1 we list the values of $w$, $\tau$, $f$, and $s$ selected for our resonator model at $20°C$. Estimates for ion $w$s fall within a range expected from atomistic molecular dynamics simulations (Delemotte et al., 2015; Lenaeus et al., 2017), and our constant ion gating relaxation times resemble voltage-averaged depolarization-dependent HH values. Our membrane surface fractions occupied by sodium and potassium channels also lie within a broad range supported by experimental studies and rough approximations to cross-sectional areas for these channels (Aidley, 1998; Jiang et al., 2003; Payandeh et al., 2011; Sula et al., 2017). Table 1 also contains values of aqueous ion diffusion coefficients $D$ and bulk concentrations $c$ of sodium, potassium and chloride ions on each side of a membrane of total thickness $L = 6$ nm (Gray and Stiles, 2018; Gray et al., 2011) and capacitance per unit area $C_m = 1.0$ μF cm$^{-2}$ (Hodgkin and Huxley, 1952; Moore and Cole, 1960).

**Table 1. Parameters for permeant ions and channels in our resonator model for the axonal membrane at a notional temperature of 293 K**

Values of $D$, $c^{\text{int}}$ and $c^{\text{ext}}$ at 20°C were taken from the literature (Hodgkin, 1958; Pilson, 1998; Perram and Stiles, 2010) and values of $f, \beta w, \tau$, and $s$ were chosen by us according to the criteria outlined following Eqs (12). The external medium is regarded as surface seawater (Pilson, 1998).

| channel or ion | Na$^+$ | K$^+$ | Cl$^-$ |
|---|---|---|---|
| membrane area fraction ($10^{-5}$) | $f_{\text{Na}}$ | $f_{\text{K}}$ | $f_{\text{Cl}}$ |
| | 10 | 3.5 | 0.5 |
| diffusion coefficient ($10^{-9}$ m$^2$s$^{-1}$) | $D_{\text{Na}}$ | $D_{\text{K}}$ | $D_{\text{Cl}}$ |
| | 1.19 | 1.78 | 1.84 |
| PMF barrier (dimensionless) | $\beta w_{\text{Na}}^{\text{O}}$ | $\beta w_{\text{K}}^{\text{O}}$ | $\beta w_{\text{Cl}}$ |
| | 3.0 | 3.0 | 6.9 |
| | $\beta w_{\text{Na}}^{\text{C}}$ | $\beta w_{\text{K}}^{\text{C}}$ | |
| | 12.8 | 10.9 | |
| | $\beta w_{\text{Na}}^{\dagger\text{O}}, \beta w_{\text{Na}}^{\dagger\text{C}}$ | | |
| | $-1.7, 8$ | | |
| ion concentration (mM) | $c_{\text{Na}}^{\text{int}}$ | $c_{\text{K}}^{\text{int}}$ | $c_{\text{Cl}}^{\text{int}}$ |
| | 50 | 400 | 40 |
| | $c_{\text{Na}}^{\text{ext}}$ | $c_{\text{K}}^{\text{ext}}$ | $c_{\text{Cl}}^{\text{ext}}$ |
| | 480.6 | 10.46 | 559.4 |
| gating relaxation time (ms) | $\tau_{\text{m}}, \tau_{\text{h}}$ | $\tau_{\text{n}}$ | |
| | 0.12, 2.5 | 2 | |
| gating steepness parameter | $s_{\text{m}}, s_{\text{h}}$ | $s_{\text{n}}$ | |
| | $0.16\,(\text{mV})^{-1}, 11$ | $0.15\,(\text{mV})^{-1}$ | |

Note that both the HH model and our present theoretical model neglect small contributions to electric currents in ion channels due to the opening and closing of the gates themselves. Such gating currents (Bezanilla, 2018; Hodgkin and Huxley,1952) are associated with motions of permanent electric multipoles of the channel gates. The displacement current $\partial \boldsymbol{D}/\partial t$ used to obtain (9b) contains small contributions due to spatial translations of gating charges and spatial rotations of higher multipole moments of gating structures within channel proteins.

When we relax the space-clamp conditions that lead to a stationary membrane action potential, propagation of the action potential occurs along a particular direction of the cylindrical $z$ axis of the axon. We define the positive $z$ direction as the direction of propagation velocity $\boldsymbol{v}$, and we discuss below initial and boundary conditions on the potential $V$ and current $i$ that determine the propagation direction. Our previous transmembrane coordinate $x$ then becomes the radial coordinate $\rho$. By taking the divergence of the Maxwell equation involving curl $\boldsymbol{B}$, we obtain $\partial(\nabla.\boldsymbol{D})/\partial t + \nabla.\boldsymbol{i}_{\text{tot}} = 0$, equivalent to the continuity or conservation of charge condition. From this equation, one can derive the HH cable equation for propagating action potentials (Hodgkin and Huxley, 1952; Jack et al., 1975; Keener and Sneyd, 2009)

$$C_m \frac{\partial V(z,t)}{\partial t} = \frac{a}{2R}\frac{\partial^2 V(z,t)}{\partial z^2} - i_\rho(z,t) \quad , \tag{13}$$

a diffusion equation for the membrane depolarization $V(z,t)$, together with the radial ionic current density $i_\rho(z,t)$ due to ions passing through their respective channels acting as a forcing term. As discussed below, the current density $i_\rho$ depends on $V$ nonlinearly, as before, so that (13) is a nonlinear propagation equation for the action potential. In (13) $a$ is the radius of the cylindrical axon and its uniform internal resistivity $R$ is defined by the longitudinal component of the internal total current density $i_z(z,t) = E_z(z,t)/R$, where $E_z$ is the longitudinal component of the internal electric field. In deriving (13) it is assumed (Hodgkin and Huxley, 1952) that the resistance of the medium external to the axon is negligible compared to the resistance of the axoplasm. This assumption is valid if the external medium has a large transverse dimension compared to the axon radius $a$. It is also assumed (Scott, 1972; Scott, 1977) that $\omega \varepsilon R \ll 1$ and that $L \ll a \ll \lambda \ll (R/\mu_0 \omega)^{1/2}$, where $\varepsilon = \varepsilon_r \varepsilon_0$ is the electric permittivity of the axoplasm with $\varepsilon_r \sim 80$ its relative permittivity (dielectric constant), $\lambda$ is the length scale characterizing the dominant short wavelengths describing the spike near the peak of the propagating action potential, $\omega$ is the frequency corresponding to $\lambda$ ($\omega \lambda \sim v$) where $v$ is the speed of the action potential pulse, and $\mu_0 = 1.257 \times 10^{-6}$ kg m C$^{-2}$ is the magnetic permeability of free space. For the squid giant axon these conditions are well satisfied since $R \sim 0.35$ ohm m, $\omega \sim 1\,\text{ms}^{-1}$, $\varepsilon R \sim 0.25$ ns (Maxwell relaxation time), $L \sim 6$ nm, $a \sim 0.25$ mm, $\lambda \sim 2$ cm (see Fig. 6 below) and $(R/\mu_0 \omega)^{1/2} \sim 17$ m.

We take the initial condition on (13) to be the longitudinally uniform resting depolarization $V(z,0) = 0$. The initial gating functions for the axon are also taken to be given by their resting-state values $m(z,0) = m_{ss}(V(z,0))$, $h(z,0) = h_{ss}(m(z,0))$, and $n(z,0) = n_{ss}(V(z,0))$ where steady-state values of the gating functions are defined by (12). For longitudinal current injection through the left-hand boundary $z = 0$ of a finite uniformly cylindrical axon of length $\ell$, the simplest boundary condition consistent with Ohm's law for the axial electric field is

$$\partial V(z,t)/\partial z\,|_{z=0} = -R\,i_{z,\text{stim}}(0,t) \quad , \tag{14a}$$

where $i_{z,\text{stim}}(0,t)$ is a transient stimulating pulse of current density directed from left to right. The corresponding local boundary condition at $z = \ell$, is assumed to take the form

$$\partial V(z,t)/\partial z\,|_{z=\ell} = 0 \quad , \; t < \ell/v \quad . \tag{14b}$$

The speed $v$ of the action potential pulse is not exactly known for our model before the calculation but can be estimated from approximate formulae for the speed (Scott, 1977; Nossal and Lecar, 1991) or by using the experimental estimate, and is roughly 20 $\text{m s}^{-1}$. Many other boundary conditions could also be imposed, e.g., specifying $V(z,t)$ at $z = 0$ and at $z = \ell$.

## 2.1 Steady-state current-voltage relation

Using data in Table 1, we solve (7) and (8) for the resting potential and ion permeabilities in the resting state of the perfused axon at $20°\text{C}$. The pumps-off resting potential at $20°\text{C}$ is $V_{\text{rest}} = -67.6$ mV. This GHK resting potential is somewhat larger than the pumps-on resting potential of about $-70$ mV for a freshly dissected squid axon (Moore and Cole, 1960). As found here, the resting potential for electrodiffusion with no pumps operating is widely believed to be more positive than in the membrane resting state with electrogenic Na-K-ATPase pumps turned on (Läuger, 1991; Mullins and Noda, 1963). For ion permeabilities in the resting state we find that $P_{\text{Na}} = 3.5 \times 10^{-8} \text{ cm s}^{-1}$, $P_{\text{K}} = 9.95 \times 10^{-7} \text{ cm s}^{-1}$ and $P_{\text{Cl}} = 1.55 \times 10^{-7} \text{ cm s}^{-1}$. Our ratio $P_{\text{Na}} : P_{\text{K}} \approx 0.035$ is lower than an experimental pumps-off estimate (Baker et al., 1962) of 0.08, and our chloride ion permeability is also lower than an experimental pumps-off value of $3.0 \times 10^{-7} \text{ cm s}^{-1}$ (Inoue, 1985). In this context we note that the analytical GHK model is not only a simplified version of (1) and (2) but is also a mean-field model that should not be expected to accurately describe how particular monatomic cations or anions pass through their channels in single file (Hodgkin and Keynes, 1955; Abercrombie, 1978; Jackson, 2006; Köpfer et al., 2014; Nelson, 2011; Tolokh et al., 2006).

The GHK resting state of the axon corresponds to resting-state or fixed point conditions of our dynamical system. Our calculated pumps-off steady-state $I - V$ curve (Fig. 3) shows that this is the only fixed point. A standard stability analysis indicates that this fixed point, like that of the HH dynamical system, is stable. The experimental pumps-on steady-state $I - V$ curve of the squid axon (Hodgkin et al., 1952) also exhibits just one fixed point. We expect a small quantitative, but not qualitative difference between $I - V$ curves for the pumps-on and pumps-off cases. Here, the steady-state membrane current $I$ is related to $\dot{V}$ in the steady-current state by (see (9)) $I = i_{tot}(\infty) = i_{Na}(\infty) + i_K(\infty) + i_{Cl}(\infty) = -C_m \dot{V}(\infty)$. Thus, $I = 0$ corresponds to the first, $\dot{V} = 0$, of the four complete steady-state, or resting-state, conditions which also include $\dot{m} = 0$, $\dot{h} = 0$ and $\dot{n} = 0$. It was from these three latter conditions that Fig. 3 was generated. It shows a plot of our membrane current versus membrane depolarization that is very similar to the experimental (Hodgkin et al., 1952) pumps-on $I - V$ curve.

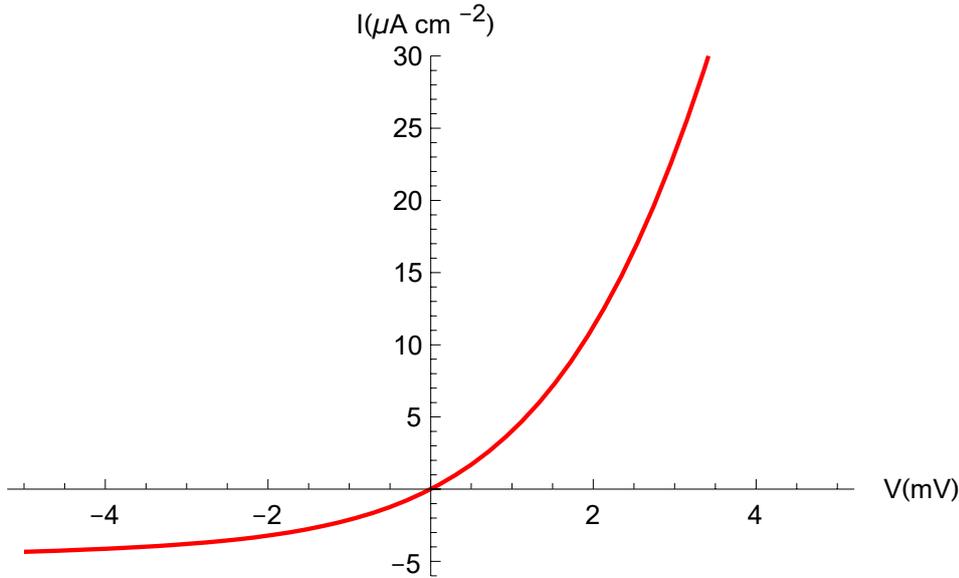

Fig. 3. Plot of the steady-state ion current $I = i_{Na}(\infty) + i_K(\infty) + i_{Cl}(\infty) = -C_m \dot{V}(\infty)$ through the membrane versus membrane depolarization $V$ showing that the origin, representing the resting state of the axon, is a fixed point. No further fixed points (where $I = 0$) become apparent when the current and voltage axes are extended.

## 2.2 Stationary membrane action potentials

After our model membrane at 20°C is stimulated by a super-threshold depolarizing voltage shock of 14 mV, we predict an action potential with a peak height of 120.3 mV some 0.41 ms later. A barely super-threshold shock of 6.551 mV produces an action potential with a peak depolarization of only 74.7 mV and a considerably longer latency of 1.95 ms. It is well-known that a more practical method for inducing a membrane action potential is to apply a brief pulse of depolarizing electric current through the membrane. In Fig. 4 we display the membrane action potential induced by a significantly super-threshold depolarizing current density pulse applied for 0.1 ms.

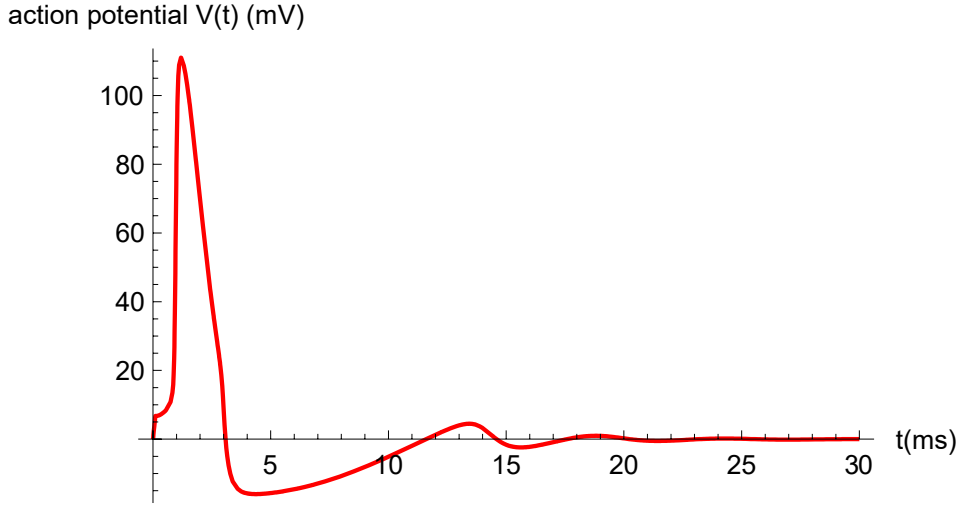

Fig. 4. Stationary membrane action potential at $20°C$ induced by a stimulating current density pulse of constant strength $-69\,\mu A\,cm^{-2}$ applied transversely for 0.1 ms at $t = 0$ ms.

According to our present model, depolarization in the action potential peaks at approximately $V = V_m^{Na} - V_{rest} = 125\,mV$, where $V_m^{Na} = 57.2\,mV$ is the membrane Nernst potential for sodium ions at $20°C$. Note from (2) or (3) that $i_{Na} = 0$ when $V_m = V_m^{Na}$ (and similarly $i_K = 0$ when $V_m = V_m^K$). The peak occurs when $t = 1.2\,ms$, some 0.2 ms after the negative sodium-channel current density has passed through its minimum value. The weaker positive potassium channel counter-current peaks close to 1.6 ms and, together with a much weaker chloride channel leakage current of the same sign, produces a small hyperpolarization or undershoot in the action potential $V(t)$. These two currents produce the well-known minimum depolarization of the action potential at about $V = V_m^K - V_{rest} = -24\,mV$ with $V_m^K = -92\,mV$, the membrane Nernst potential for potassium ions at $20°C$. Between about 4 ms and 12 ms a weakly negative ungated chloride ion current in combination with an increasingly negative depolarizing sodium current take the action potential back through its resting-state value. After this recrossing of the resting state, the small depolarizing sodium current trough followed by a slightly larger secondary hyperpolarizing potassium current peak, lead to resonant (oscillatory) decay in the tail of the action potential.

We discussed earlier (see Fig. 1B) the corresponding membrane action potential predicted by our model in response to a brief hyperpolarizing current stimulus.

## 2.3 Stationary membrane spike trains associated with reduced concentrations of external $Ca^{2+}$

Lowering the small concentration of divalent cations such as external $Ca^{2+}$ enhances the excitability of the plasma membrane. Repetitive firing of action potentials then occurs (Frankenhaeuser and Hodgkin, 1957; Guttman and Barnhill, 1970; Guttman et al., 1980). We regard both the free energy barrier $w_{Na}^O$ of a sodium ion traversing the open sodium activation gate and a constant transverse stimulating current density $i_{stim}$ (applied to the membrane for 0.1 ms) as bifurcation control parameters. Our present electrodiffusion model then predicts repetitive firing.

Consider first the external calcium ion concentration taking its normal seawater value of 10.52 mM (Pilson, 1998). When we then set $i_{stim} = -65\,\mu A\,cm^{-2}$ and let $\beta w_{Na}^O = 3.0$, as in Table 1, the stimulating current is subthreshold so no action potential occurs. We now lower the control parameter $i_{stim}$ from $-65\,\mu A\,cm^{-2}$ to the superthreshold value $-69\,\mu A\,cm^{-2}$ for 0.1 ms and maintain $\beta w_{Na}^O$ at the value 3.0. A codimension-1 bifurcation then can be induced and generates the standard isolated action potential illustrated in Fig. 4. Calcium ions in normal seawater, at a concentration of 10.52 mM, bind to negative sites in the vestibules of sodium ion channels (Armstrong, 1999) of a squid axon. Such reversible binding obstructs the free passage of sodium ions from the external electrolyte into these channels. These calcium ions then repel incoming sodium ions through both short-range 'steric hindrance' and longer-range Coulombic forces. These forces contribute to the free energy barrier $\beta w_{Na}^O = 3.0$. Lowering calcium ion concentrations in the external electrolyte reduces the local concentrations of these divalent cations inside vestibules of sodium channels and thereby lowers the free energy barrier. We next keep the brief stimulating current density $i_{stim}$ at the depolarizing value of $-69\,\mu A\,cm^{-2}$ and reduce the local free energy barrier $\beta w_{Na}^O$ from 3.0 to 1.48. We find, from our simple model that the excitability of the axonal membrane then increases to the extent that firing of repetitive action potentials, depicted in Fig. 5, commences. The transition of the spiking pattern from the isolated spike of Fig. 4 to the spike-train of Fig. 5 below implies that the squid axon responds with a codimension-2 bifurcation to the combined effects of the decrease in $\beta w_{Na}^O$ together with the usual depolarizing influence of the stimulating current-density.

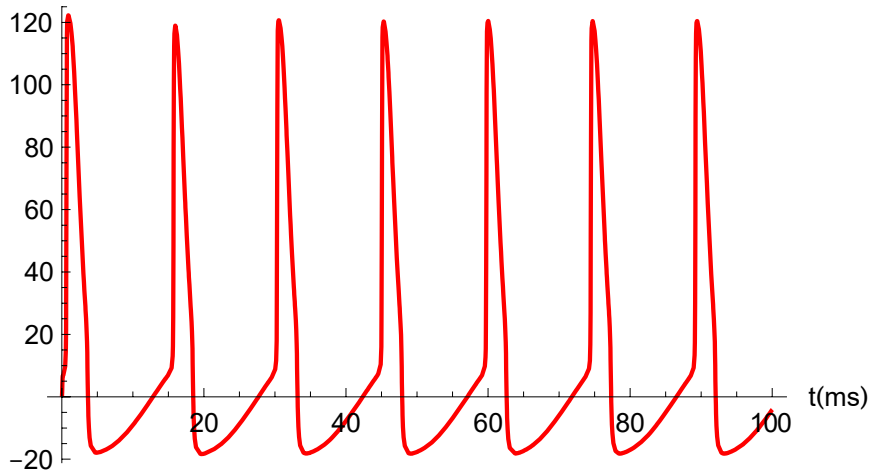

Fig. 5. Persistent spike train of membrane action potentials, triggered by a 0.1 ms pulse of transverse depolarizing membrane current-density $i_{stim} = -69\,\mu A\,cm^{-2}$ when low external calcium ion concentrations in the bulk electrolyte reduce the free energy barrier $\beta w_{Na}^O$ of sodium ions from 3.0 to 1.48 inside open sodium channels of the perfused squid axon at 20°C. Amplitudes and frequencies of action potentials in this effectively infinite spike train

are predicted to be rather insensitive to potential barriers in the range $0 < \beta w_{Na}^O < 1.48$ and the spiking frequency possibly reflects a typical refractory period for rebound action potentials with sodium activation barriers somehat lower than in normal surface seawater.

The bifurcation from firing of isolated action potentials to repetitive firing of action potentials can be regarded as an interesting case of rebound firing. As the height $\beta w_{Na}^O$ of the sodium activation barrier decreases from 3.0 to 1.48, the maximum hyperpolarization in the undershoot following the generation of an isolated action potential increases (due to increased $K^+$ and $Cl^-$ repolarization currents following the larger $Na^+$ generated depolarization peak) and the rebound depolarization subsequently overshoots the resting potential after 12.81 ms. At this second resting potential the total ionic current density, dominated by the sodium ion influx, exceeds the depolarization threshold for the formation of a second membrane action potential so firing of a second action potential occurs, as indicated in Fig. 5. This rebound action potential in turn, generates yet another rebound action potential, and thus, repetitive firing ensues. According to our model, repetitive rebound spiking is also induced by increasing gating relaxation times, even in the absence of external calcium ions. An alternative way to account for enhanced $Na^+$ current leading to rebound firing is as follows: we keep the current density $i_{stim}$ of the brief stimulating current at the depolarizing value of $-69$ $\mu A\,cm^{-2}$, maintain the local free energy barrier $\beta w_{Na}^O$ at the value 3.0, but reduce the steepness parameter $s_m$ in (12a) for the voltage dependence of the sodium $m$-gating from 0.16 to 0.14 $(mV)^{-1}$. The tail of primary action potential then rebounds more rapidly and overshoots the resting potential after only 11.56 ms. At this juncture the total ionic current density, again dominated by the sodium ion influx, exceeds the depolarization threshold for the generation of a secondary stationary membrane action potential. This argument for the formation of a second action potential, using the steepness parameter $s_m$ as a codimension-2 bifurcation parameter in place of the sodium-channel activation barrier $\beta w_{Na}^O$, can be repeated to account for periodic rebound firing, similar to that depicted in Fig. 5, but with a slightly higher frequency of firing as can be inferred from comparing the resting-potential overshoot times (11.56 ms versus 12.81 ms) for the respective spike trains.

## 2.4 Propagating action potentials

Symbolic, numerical and graphical capabilities of Wolfram's Mathematica software permit the standard HH cable equation (13) in conjunction with our gating equations

$$\frac{\partial m(z,t)}{\partial t} = \frac{m_{ss}(V(z,t)) - m(z,t)}{\tau_m} \quad , \tag{15a}$$

$$\frac{\partial h(z,t)}{\partial t} = \frac{h_{ss}(m(z,t)) - h(z,t)}{\tau_h} \quad , \tag{15b}$$

$$\frac{\partial n(z,t)}{\partial t} = \frac{n_{ss}(V(z,t)) - n(z,t)}{\tau_n} \quad , \tag{15c}$$

and (3)-(6) to be solved routinely by using the numerical method of lines. The steady-state gating functions are the expected generalizations of (12) with $V \equiv V(z,t)$, $m \equiv m(z,t)$, $h \equiv h(z,t)$ and $n \equiv n(z,t)$. Here, $V(z,0) = 0$ and the initial values of the gating functions are $m(z,0) = m_{ss}(V(z,0)) = m_{rest}$, $h(z,0) = h_{ss}(m(z,0)) = h_{rest}$ and $n(z,0) = n_{ss}(V(z,0)) = n_{rest}$. Because $m(z,t)$ etc are determined by $V(z,t)$, for which boundary conditions have been specified, independent boundary conditions for $m(z,t)$ etc cannot also be imposed. To estimate propagation speeds, we have considered a 50 cm section of a squid giant axon with an internal longitudinal resistivity of $R = 35.4$ ohm cm and a longitudinally constant radius of 0.238 mm (Hodgkin and Huxley, 1952). The initial condition corresponds to the resting axon as described in Table 1 with all transmembrane pumps switched off. A propagating impulse can then be triggered by the boundary condition (14a) at $z = 0$ by applying a constant longitudinal current density pulse $i_{z,stim}$ of 7.3 A m$^{-2}$ for 0.50 ms at $t = 0.01$ ms. The influence of such a stimulating current on the axonal membrane is depolarizing for a small region near $z = 0$ just after the stimulus ceases. The second boundary condition (14b) stipulates zero longitudinal electric current density at $z = 50$ cm. A graphical representation of $V(z,t)$, from the solution to these equations, is presented in Fig. 6. Note that the condition $t < \ell/v$ in (14b) is satisfied for the pulse shown in Fig. 6 since $\ell/v$ is about 25 ms.

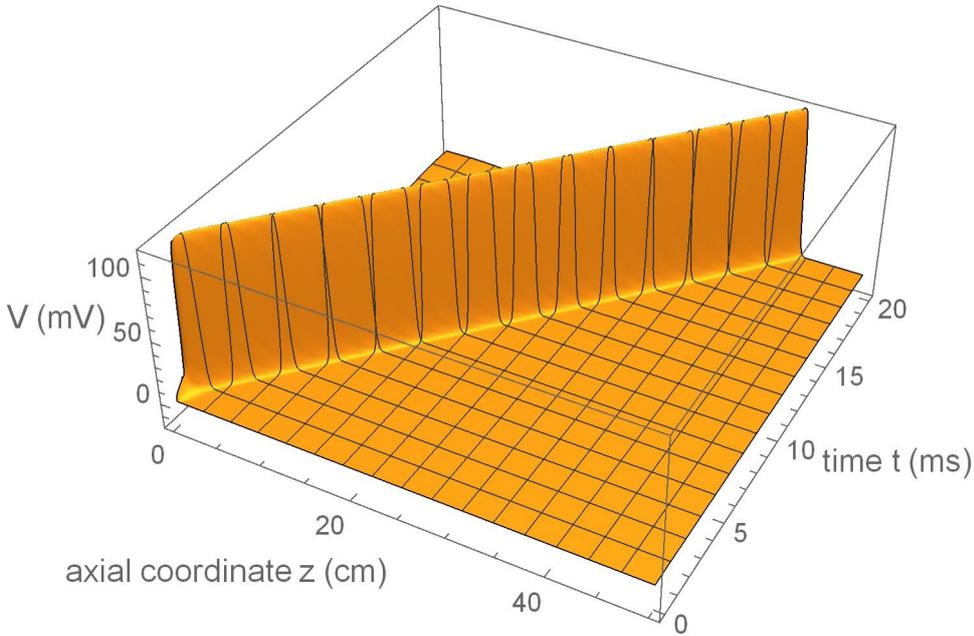

Fig. 6. Plot of an action potential $V(z,t)$ of peak height 119.5 mV propagating at 20°C from left to right along a 50 cm section of squid axon induced by a stimulating current pulse $i_{z,stim}(z = 0,t) = 7.3$ A m$^{-2}$ between $t = 0.01$ ms and $t = 0.51$ ms. From the analysis of this result, we find a propagation speed of $22.3\,\mathrm{m\,s^{-1}}$ at 20°C, in good agreement with the estimated experimental speed (Hodgkin and Huxley, 1952) of $21.2\,\mathrm{m\,s^{-1}}$ at 18.5°C.

Jack et al. (1975) pointed out that the speed of a nerve impulse is faster close to the stimulated end of a finite axon with a uniform diameter. For a long axon, this speed rapidly becomes spatially homogeneous as we move away from the boundaries. We estimate the propagation speed of $22.3\,\mathrm{m\,s^{-1}}$, indicated in Fig. 6, from the time taken for the peak of our action potential to travel from $z = 24.95\,\mathrm{cm}$ to $z = 25.05\,\mathrm{cm}$.

    Both the direction and duration of our constant stimulating longitudinal current required to elicit right-propagating action potentials are important. As shown in Fig. 6, initiation of a normal propagating action potential by a positive electric current density applied for 0.5 ms at $z = 0$ usually requires a superthreshold longitudinal current density with a constant magnitude of at least $7.3\,\mathrm{A\,m^{-2}}$. This positive electric current density, associated with cations moving from left to right or with anions moving from right to left, depolarizes the axonal membrane, initially at $z = 0$. This depolarization later spreads to values of $z > 0$. When the magnitude of this stimulating current density $i_{z,\mathrm{stim}}(0,t)$ is sufficiently large, transient subthreshold oscillations resonate and a normal action potential fires (Izhikevich, 2001). If, on the other hand, we apply a much stronger negative longitudinal current-density pulse $i_{z,\mathrm{stim}}(0,t) = -68\,\mathrm{A\,m^{-2}}$ to the squid axon at $z = 0$ between $t = 0.01\,\mathrm{ms}$ and $t = 0.51\,\mathrm{ms}$ the left end of the neural membrane of the squid axon first hyperpolarizes. As illustrated in Fig. 7 below, the axon then depolarizes rapidly, resonates and fires (Izhikevich, 2001) some 9 ms after the stimulating current pulse ceases. An inhibitory rebound propagating action potential, with a profile similar to that of the space-clamped action potential, displayed in Fig.1B, then propagates along the positive $z$ direction. Speeds of regular and rebound modes are found to be very similar, as expected physically, but the error in the speed of our rebound mode is somewhat larger than that for the regular mode. As noted earlier, we are unaware of experimental investigations of rebound propagating action potentials induced solely by brief hyperpolarizing axial currents in giant axons of squid.

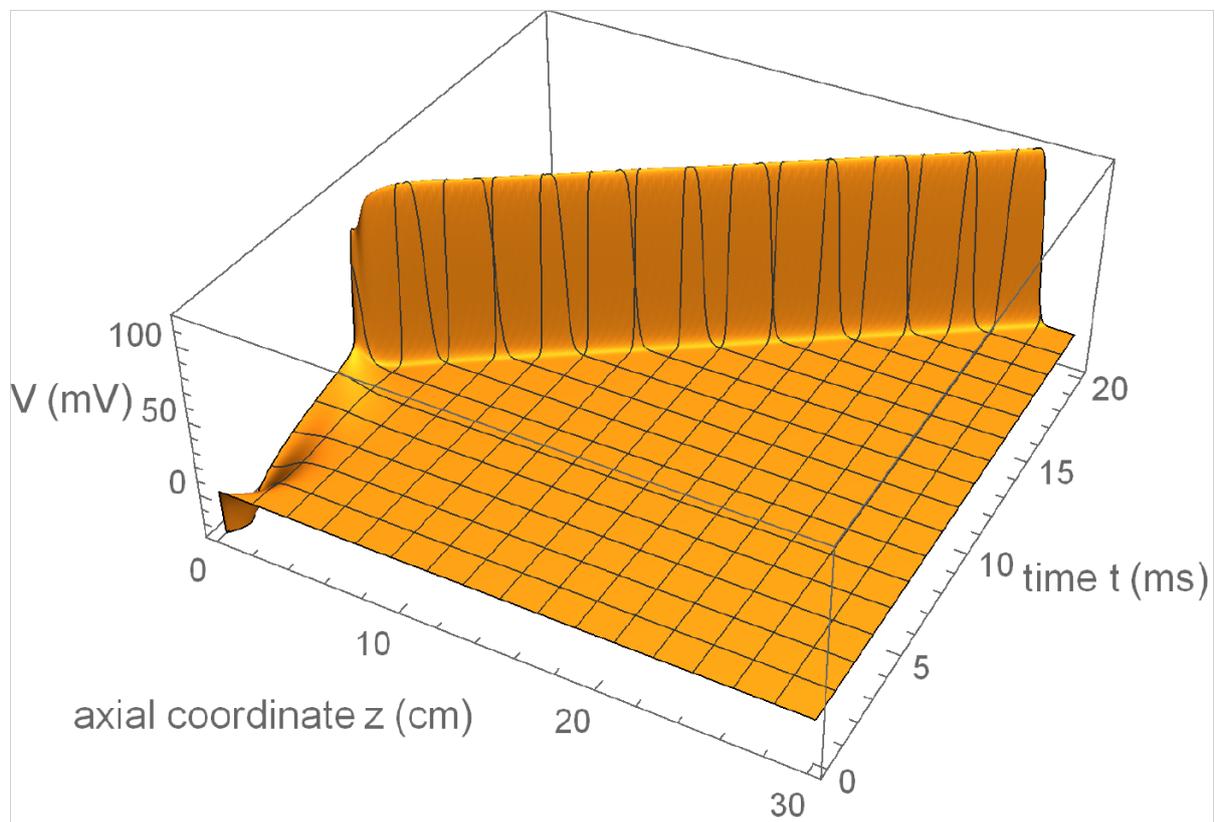

Fig. 7. Here, the left-hand side boundary condition at $z=0$ is a $-68\,\text{A}\,\text{m}^{-2}$ stimulating longitudinal current density, directed from right to left, and applied for 0.5 ms. Our electrodiffusion model then predicts an initial hyperpolarization of the left end, followed by an action potential with a latency of about 9 ms. The speed of the left-to-right propagating rebound action potential, with a peak depolarization of 119.5 mV, is 22 $\text{m}\,\text{s}^{-1}$ at $20°\text{C}$.

When the squid axon is bathed in surface seawater with a sufficiently reduced external calcium ion concentration, our model predicts that multiple travelling waves (spike trains) can be induced by either positive or negative longitudinal current stimuli. This can be achieved by lowering either the sodium activation barrier $\beta w_{\text{Na}}^{\text{O}}$ or the steepness of the sodium activation gating parameter $s_{\text{m}}$, as in Section 2.3 dealing with conditions sufficient for formation of the single spike train illustrated in Fig. 5.

## 3. Concluding remarks

We have extended the quantitatively crude but simple and physically-based nonlinear GHK model, for steady-state electric currents and resting potentials in the giant axon of the squid, into the time domain. We can then describe both stationary space-fixed and propagating action potentials more naturally by avoiding speculative gating-state dependencies, such as $m^3 h$ and $n^4$, introduced by HH to quantify ionic current densities in voltage-gated sodium and potassium channels. Our very simple theoretical model provides a unified nonlinear treatment of both resting and action potentials in perfused (Baker et al., 1961,1962) giant axons of squid. In these perfused axons the intact axoplasm of a live squid has been replaced by its original contents minus the ATP required to power various pumps. Voltage-dependent relaxation times for channel-gating, required by the HH model, are no longer needed. This model can be extended to describe resting and action potentials for axons with pumps on. To model a live squid, the external electrolyte should also use the composition of cephalopod blood in which the external potassium ion concentration is roughly double that of seawater (Hodgkin, 1958).

The chloride ion concentration within the giant axon requires careful consideration. Steinbach (1941) originally measured a concentration of 40 mM, subsequently endorsed by Hodgkin (1958). This internal chloride ion concentration is very similar to that expected if the chloride ions are passively distributed, in compliance with the Nernst equation, across the squid axolemma. Subsequent experimental investigations, however, by Koechlin (1955) and by Keynes (1963) yielded estimates of about 140 mM and 123 mM, respectively, that have been supported by the subsequent discovery of the ATP-linked sodium- potassium-chloride cotransporter (Russell, 2000) which functions as a chloride ion pump. When our perfused squid axon loses its ATP, both the Na-K-ATPase pumps and the chloride ion cotransporter pumps fail to operate. The internal chloride ion concentration of 40 mM recommended by Hodgkin (1958) therefore seems most appropriate for our present calculations on electrodiffusion. We then predict a resting potential $V_{\text{m}}$ of about $-68$ mV and a membrane action potential with peak depolarization close to 120 mV at $20°\text{C}$. A predicted speed for a propagating action potential close to the experimental estimate of HH on a non-perfused squid axon at $18.5°\text{C}$ is then obtained.

Unlike its HH (pumps-on) counterpart, our new theoretical pumps-off model, in agreement with experiments (Clay, 1998) on perfused axons does not predict spike trains of repetitive action potentials during prolonged constant-current stimulation of any size. On the other hand, it does predict, in agreement with experiment, spike-train formation, more simply and physically than previous simulations (Guttman et al., 1980), following brief current stimulation of axons bathed in low concentrations of calcium ions.

Rebound spiking following prolonged hyperpolarizing stimulation found at 18.5°C by HH, but not predicted at this temperature by the HH model, is predicted by our present model. As shown in Fig.1B, rebound spiking triggered by brief (0.1 ms) hyperpolarization is also predicted at room temperature by our present model, but only at much lower temperatures (*ca.* 6 °C) by the HH model. Such stationary membrane action potentials and their propagating counterparts, due to brief hyperpolarizing stimulation at room temperature, do not yet appear to have been sought experimentally in giant axons.

Temperature dependences of resting and action potentials warrant discussion, both for their intrinsic interest and because changes in model parameters induced by temperature changes can lead to bifurcations in action potential behavior as discussed toward the end of this section. HH ascribe temperature dependence of action potentials entirely to channel-gating relaxation times. Many parameters in our own model depend on temperature and therefore modify the membrane potential. In the original HH model, $T$ dependence of the action potential arises through $T$ dependence of forward and reverse rate coefficients such as $\alpha_n$ and $\beta_n$ for a typical potassium-channel gating transition Closed $\Leftrightarrow$ Open. These rate coefficients are related to the relaxation time and steady-state gating value. For potassium channel activation $\tau_n = 1/(\alpha_n + \beta_n)$ and $n_\infty = \alpha_n/(\alpha_n + \beta_n)$. Thus, the HH rate coefficients $\alpha_n$ and $\beta_n$ for potassium-ion gating can, in principle, be determined by fitting to voltage- or patch-clamp data (Forrest, 2014; Tiwari and Sikdar, 1999) at various temperatures. HH partially carry out the former procedure, but (over)simplify it by (tacitly) assuming that forward and reverse rate constants have exactly the same temperature dependences; the assumption that gating barriers for forward and reverse gating transitions are identical is not justified by HH. Together with the additional simplifying assumption that temperature dependences of the rate coefficients $\alpha(V,T)$ and $\beta(V,T)$ are the same for all sodium- and potassium-channel gates in squid, they find that their gating relaxation times $\tau$ scale exponentially with $T$, but assume their voltage-clamped steady-state gating variables $m_\infty$, $h_\infty$ and $n_\infty$ are independent of $T$. Physically, we expect $m_\infty$, $h_\infty$ and $n_\infty$ to depend strongly on $T$ except for large and small values of $V$ where they have little freedom to assume other than their limiting values of zero for a closed gate and unity for an open gate. With a reference temperature chosen to be 6.3°C, the common HH scaling factor for gating relaxation times $\tau$ is $1/\phi(T)$ in which $\phi(T) = 3^{(T-6.3)/10}$ for every gate within sodium and potassium channels. A simplified HH study of a non-squid potassium channel (Forrest, 2014), using rate constants $\alpha_n(V,T)$ and $\beta_n(V,T)$, suggests that not only is $n_\infty$ temperature dependent, but also that commonly assumed temperature independence of activation barriers in the transition-state-theory Arrhenius forms (Tsien and Noble, 1969) for $\alpha_n$ and $\beta_n$ may not be justified. Temperature affects gating transition rates not only directly through gating molecular dynamics, but possibly also indirectly through its effect on membrane fluidity (Rosen, 2001). As discussed next, other factors determining membrane potentials, such as open- and closed-channel permeabilities, are also temperature sensitive and have been studied experimentally (e.g., Pahapill and Schlichter, 1990).

The reference temperature for our present model is 20°C. As seen from the results (9b), (13), (3) and (7), the dependence $V(t,T)$ of membrane depolarization on temperature $T$ arises mainly through: the relatively weak dependence on membrane capacitance $C_m$, the somewhat stronger dependence on axoplasm resistivity $R$, and the larger dependence on the various factors determining the channel currents (3). There are explicit $T$ (or $\beta = 1/k_B T$) factors in (3), as well as strong implicit $T$ dependences via the membrane potentials $V_{rest}$ and $V(t)$ in $V_m(t)$, in the Nernst potentials $V_m^{Na}$ and $V_m^K$ which largely determine the peak height and trough depth of an action potential impulse, and also in the ionic permeabilities $P = (fD/L)\exp(-\beta w)$. Thus the friction on a permeating ion, expressed by a channel diffusion coefficient $D$, depends strongly on $T$, as does our instantaneous PMF $w = w(t)$, expressed for $K^+$ ions, for example, by (5). A PMF $w$ depends relatively weakly on $T$ through the potential barriers $w^O$ and $w^C$, but strongly through the channel-gating variable $n$ for a $K^+$ channel, or through both $m$ and $h$ for a $Na^+$ channel. Our channel-gating variables $m$, $h$, and $n$ are temperature dependent through both channel-gating relaxation times $\tau(T)$ and also through steady-state values $m_{ss}(V,T)$, $h_{ss}(V,T)$, and $n_{ss}(V,T)$ of these variables. Dependences of these steady-state gating variables on $V$ and $T$ are defined by (12).

Oscillations in the decay of an action potential or a subthreshold $V(t)$ arise, not as originally assumed by Cole (1941) via explicit circuit inductance, but more naturally through delays in ion current, or *effective* inductance, associated with moderately large values of $\tau_n$ and $\tau_h$ (Hodgkin and Huxley,1952; Chandler et al.,1962). HH discussed how "inertia" in the $K^+$ current arises from the relaxation time for potassium-channel gating, and Chandler *et al.* derive the explicit relation between the effective inductance $L_n$ and $\tau_n$ for a $K^+$ current (with similar relations for inductances associated with the other gates). Both groups used a linearized version of the four nonlinear HH dynamical equations involving $\dot{V}, \dot{m}, \dot{h}, \dot{n}$, valid near the fixed point, i.e., for small subthreshold potentials or in the far tail of an action potential pulse. A similar analysis can be carried out for our new model by linearizing the four dynamical equations (9b), (11a,b,c) of our model. Under typical physiological conditions both our new model and the HH model are resonators rather than integrators. For our new model, oscillations in the tail of an action potential pulse are clearly evident in Figs. 4 and 1B. Oscillations in the HH model are smaller, but can be seen in the enlarged-scale Fig. 24 of the original HH model, or if we enlarge the scale of Fig.1A for times beyond about 25 ms. According to the HH model a modest 10 degree decrease in temperature from 16.3°C to 6.3°C increases all $\tau$ values and all effective inductances by a factor of 3 and causes decay of the undershoot in an isolated action potential to exhibit enhanced subthreshold oscillations. Our present model suggests that not only does a small decrease in temperature enhance oscillatory/resonator behavior, but also that if all our gating relaxation times increase by a factor as small as 1.4, a bifurcation to repetitive oscillatory or tonic firing similar to that illustrated in Fig. 5 occurs. Even if the potassium gating relaxation time $\tau_n$ alone increases from 2 ms to 2.4 ms, our model predicts a bifurcation to repetitive or tonic firing of action potentials. Previous studies based on the HH model (both full and reduced versions) have examined bifurcation behavior due to changes in parameters such as temperature, stimulating current, sodium permeability (linked to external $Ca^{2+}$ concentration, as discussed in Section 2.3), potassium permeability, and external potassium concentration (Ermentrout and Terman, 2010; Feudel et al., 2000; Guttman and Barnhill, 1970; Hahn and Durand, 2001; Aihara and Matsumoto, 1987; Hodgkin and Katz, 1949b; Horng and Huang, 2006; Izhikevich, 2007; Rubin and Wechselberger, 2007; Schmid et al., 2004). Fixed-point to limit-cycle bifurcations, and bifurcations such as transitions to bistability, bursts (individual and periodic), and other nonperiodic behaviours such as resonator to integrator transition, and transitions to chaos, have been found. We plan to explore additional bifurcations for our new model (both full and reduced), and to compare the results with predictions from the HH model and with experimental data.

Far more elaborate molecular dynamics models, incorporating detailed kinetics of ATPase ion pumps, detailed structural features of ion channels and their gating regions, more realistic spatiotemporal electrolyte distributions within ion channels and extending beyond their orifices for an *in-vivo* axonal membrane, can be envisaged. Our dynamical generalization of the well-known electrodiffusion-based GHK model for resting potentials permits significant mechanistic rationalization of the phenomenological conductance-based HH model for action potentials in a perfused squid axon. It captures, very simply, the nonlinear physics essential for understanding both resting and action potentials in giant axons of squid.